# Unconventional bias-dependent tunneling magnetoresistance in van der Waals ferromagnetic/semiconductor heterojunctions

Wenkai Zhu,[1,†] Hui Wen,[1,2,†] Shouguo Zhu,[1,2,†] Qirui Cui,[3,†] Shihong Xie,[1,4] Meng Ye,[1] Gaojie Zhang,[5,6] Hao Wu,[5,6] Xiaomin Zhang,[1,2] Weihao Li,[1,2] Yuqing Huang,[1] Jing Zhang,[1] Lixia Zhao,[7] Amalia Patanè,[4] Haixin Chang,[5,6,*] Lin-Wang Wang,[1] Kaiyou Wang[1,2,*]

[1]State Key Laboratory of Superlattices and Microstructures, Institute of Semiconductors, Chinese Academy of Sciences, Beijing 100083, China

[2]Center of Materials Science and Optoelectronics Engineering, University of Chinese Academy of Sciences, Beijing 100049, China

[3]Department of Applied Physics, School of Engineering Sciences, KTH Royal Institute of Technology, AlbaNova University Center, Stockholm SE-10691, Sweden

[4]School of Physics and Astronomy, University of Nottingham, Nottingham NG7 2RD, United Kingdom

[5]Center for Joining and Electronic Packaging, State Key Laboratory of Material Processing and Die & Mold Technology, School of Materials Science and Engineering, Huazhong University of Science and Technology, Wuhan 430074, China

[6]Wuhan National High Magnetic Field Center, Huazhong University of Science and Technology, Wuhan 430074, China

[7]School of Electrical and Electronic Engineering, Tiangong University, Tianjin 300387, China

[†]These authors contributed equally to this work

*Corresponding author. E-mail: hxchang@hust.edu.cn; kywang@semi.ac.cn

**Abstract**

Two-dimensional van der Waals (vdW) ferromagnetic/semiconductor heterojunctions represent an ideal platform for studying and exploiting tunneling magnetoresistance (TMR) effects due to the versatile band structure of semiconductors and their high-quality interfaces. In the all-vdW magnetic tunnel junction (MTJ) devices, both the magnitude and sign of the TMR can be tuned by an applied voltage. Typically, as the bias voltage increases, first the amplitude of the TMR decreases, then the sign of the TMR reverses and/or oscillates. Here, we report on an unconventional bias-dependent TMR in the all-vdW $Fe_3GaTe_2$/GaSe/$Fe_3GaTe_2$ MTJs, where the TMR first increases, then decreases, and finally undergoes a sign reversal as the bias voltage increases. This dependence cannot be explained by traditional models of MTJs. We propose an in-plane electron momentum ($k_\parallel$) resolved tunneling model that considers both the coherent degree of $k_\parallel$ and the decay of the electron wave function through the semiconductor spacer layer. This can explain well the conventional and unconventional bias-dependent TMR. Our results thus provide a deeper understanding of the bias-dependent spin-transport in semiconductor-based MTJs and offer new insights into semiconductor spintronics.

1. **Introduction**

   Semiconductor spintronics, with its primary focus on spin injection, manipulation, and detection in semiconducting materials, offers new avenues for designing novel spin storage, spin logic, spin photodetection and light-emitting devices[1-3]. The discovery of the giant magnetoresistance (GMR) effect in ferromagnet/non-magnet/ferromagnet full-metal heterostructure spin-valve devices, caused by spin-scattering mechanisms, has ignited a surge of interest in spintronic research[4-6]. Subsequently, improved tunneling magnetoresistance (TMR) effects were achieved in magnetic tunneling junction (MTJ) devices based on insulating spacer layers, which mainly rely on spin tunneling mechanisms[7,8]. Numerous studies have shown that the magnitude of TMR is not only related to the properties of the ferromagnetic metal electrodes, but is also closely associated with the band structure of the barrier layer[9,10]. Compared with metallic and insulating spacer layers, a semiconducting spacer layer represents a versatile tool for tuning the electron transmission in an MTJ by changing the tunneling barrier height as well as the band structure[11]. However, the efficiency of spin injection in traditional covalently bonded ferromagnetic/semiconductor heterojunction devices is limited by the lattice mismatch and conductivity mismatch between semiconductors and ferromagnetic metals.

Combining novel two-dimensional (2D) van der Waals (vdW) layered ferromagnetic metals with 2D semiconductors via van der Waals forces enables the creation of all-vdW MTJs devoid of lattice mismatch concerns[12-15]. This produces atomically sharp interfaces and reduces hybridization, thus enhancing electron tunneling and overcoming the conductivity mismatch. Large TMRs of 300% and 120% at low temperatures have been observed in the all-vdW $Fe_3GeTe_2$/hBN/$Fe_3GeTe_2$ and $Fe_3GeTe_2$/$WSe_2$/$Fe_3GeTe_2$ MTJs, respectively[16,17]. Also, a large room-temperature TMR of 85% has been obtained in the all-vdW $Fe_3GaTe_2$/$WSe_2$/$Fe_3GaTe_2$ MTJs, which is much larger than that achieved in traditional covalently bonded ferromagnetic/semiconductor heterojunctions[18,19]. Interestingly, all-vdW semiconductor-based MTJs show a strong bias-dependent spin polarization and thus TMR [16-24]: the magnitude of the TMR tends to decrease with increasing bias and a sign reversal can occur at large biases. By focusing solely on the electron spin states within the metallic ferromagnet involved in the tunneling, the initial decrease and subsequent sign reversal of TMR with increasing bias voltage can be understood[16,22,23].

In this work, we report on a different bias dependence of the TMR in all-vdW MTJs that employ the room temperature perpendicular ferromagnet $Fe_3GaTe_2$ as the spin injection and detection electrodes, and the semiconductor GaSe as the tunnel barrier. When a bias of 0.5 V is applied between the two electrodes, we observe a maximum TMR of 107% at 10 K and of 25% at room-temperature. In contrast to the previously reported bias-dependent TMR in all-vdW MTJs, an unconventional, M-shaped, bias-dependent TMR behavior is observed, which cannot be explained using tunneling models in the literature[16,21,22]. To elucidate the experimental observations, we propose an in-plane momentum $(k_\parallel)$ resolved tunneling model, where both the electron wave function decay and the coherent degree of $k_\parallel$ through the GaSe spacer layer are taken into consideration. Using this model, we can explain not only the unconventional bias-dependent TMR in $Fe_3GaTe_2/GaSe/Fe_3GaTe_2$, but also the conventional bias-dependent TMR in $Fe_3GeTe_2/GaSe/Fe_3GeTe_2$. Our work provides a deep understanding of the bias-dependent TMR in semiconductor-based MTJs, offering a route to the optimization of high-performance all-vdW MTJs.

## 2. Results and Discussion

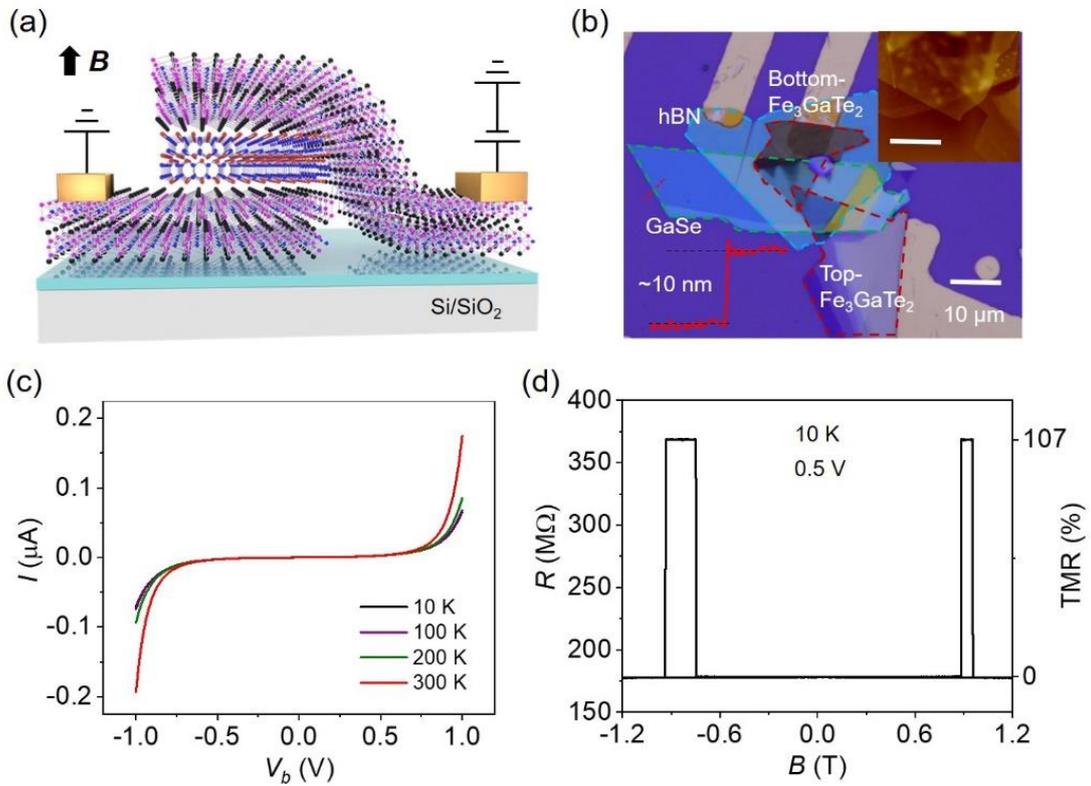

Fig. 1. Device structure and TMR effect. (a) Schematic diagram illustrating the $Fe_3GaTe_2/GaSe/Fe_3GaTe_2$ MTJ device structure and experimental setup. (b) Optical image of the device, where the different vdW nanoflakes are outlined with dashed lines of different colors. AFM measurements indicate that the GaSe layer has a thickness of 10 nm. (c) Temperature-dependent $I$-$V_b$ curves measured in the parallel state. (d) Resistance hysteresis loop of the device with the bias voltage of 0.5 V at 10 K, indicating a TMR of 107%.

The all-vdW Fe$_3$GaTe$_2$/GaSe/Fe$_3$GaTe$_2$ MTJ devices were fabricated by using a dry-transfer technique (details in the Experimental Section). The schematic diagram of the device structure and measurement setup is shown in Fig. 1a. The optical image of the device is shown in Fig. 1b, where the different vdW nanoflakes are outlined with dashed lines of different colors. The atomic force microscope (AFM) measurements indicate that the thickness of the GaSe spacer layer is 10 nm. We examine two additional devices with a GaSe layer thickness of 5 nm and 8 nm, as shown in the Supplementary Information. The top and bottom Fe$_3$GaTe$_2$ have different thicknesses (approximately 20 and 10 nm, respectively) to distinguish their coercivities. The nonlinear current-voltage ($I$-$V_b$) curves of the device are shown in Fig. 1c at different temperatures in parallel magnetic states. The weak temperature dependence of the current indicates that the transport is dominated by tunneling. By applying a constant bias voltage of 0.5 V across the device at 10 K, as shown in Fig. 1d, the resistance was monitored while scanning the out-of-plane external magnetic field ***B***. The resistance of the device in the parallel (P) and antiparallel (AP) magnetic states of the two ferromagnet layers is 178.1 MΩ and 369.1 MΩ, respectively. This corresponds to a TMR $= (R_{AP} - R_P)/R_P = I_P/I_{AP} - 1 = 107\%$ at 10 K, where ($R_P$, $I_P$) and ($R_{AP}$, $I_{AP}$) represent the resistance and measured current values of the device under the P and AP states, respectively. The TMR of the device under the fixed voltage of 0.5 V decreases with increasing temperature, going to zero at 400 K (Fig. S1), which is due to the decrease in the spin polarization of the ferromagnet. The TMR reaches a value of 25% at room temperature.

We studied the TMR effect of the device under different positive and negative biases, as shown in Figs. 2a-b. The TMR values under different biases are presented in Fig. 2d as open squares. Furthermore, we separately measured the $I$-$V_b$ curves of the device under P and AP states (Fig. 2c), from which we extracted the TMR as a function of the bias voltage (Fig. 2d), which match well with those obtained from Figs. 2a-b. The TMR of the device first increases and then decreases with the bias voltage, reaching its maximum value at a bias of 0.5 V. In the two additional devices with different GaSe thicknesses, we also observed a similar M-shaped bias-dependent behavior, verifying the reproducibility of the phenomenon (Figs. S2-4). Figure 2d shows the bias-dependent TMR curves of the device at several specific temperatures. The M-shape bias-dependent TMR behavior is robust with temperature, with the maximum TMR values around 0.5 V, indicating a stable tunneling mechanism as the origin of the TMR. This M-shaped bias-dependent TMR phenomenon is qualitatively different from that reported for $Fe_3GeTe_2$ and $Fe_3GaTe_2$ based MTJs, where the magnitude of the TMR decreases with increasing the bias voltage[16-20,22,24,25]. We examine this anomalous behavior and discuss differences amongst different MTJ devices.

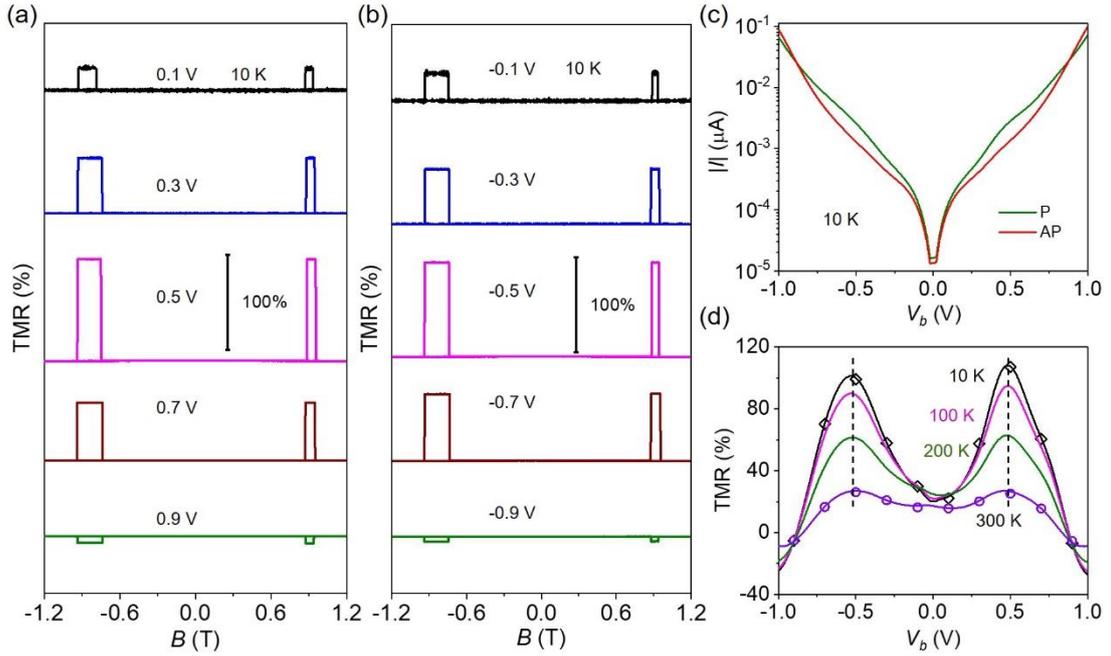

Fig. 2. Bias-dependent TMR. (a) TMR curves under positive bias in the range of 0.1 to 0.9 V at 10 K. (b) TMR curves under negative bias in the range of -0.1 to -0.9 V at 10 K. (c) The $I$-$V_b$ curves of the device in parallel and antiparallel states at 10 K. (d) TMR as a function of the bias voltage at several specific temperatures.

In previous elastic tunneling models, the band structure of the spacer layer and the scatterings of the in-plane momentum $k_\parallel$ have not been taken into account, which can explain the conventional bias dependent TMR[22]. However, the observed M-shape bias-dependent TMR behavior here can not be explained by using these existing models. To our knowledge, when an electron tunnels through a thin semiconductor barrier layer, its wave function decay is strongly dependent on $k_\parallel$, due to the band structure of the barrier layer. Considering the similar crystal structures and relatively small lattice mismatch of Fe₃GaTe₂ and GaSe (details see Fig. S5), we utilize a $k_\parallel$-resolved tunneling model for the TMR calculation within the Fe₃GaTe₂/GaSe/Fe₃GaTe₂ MTJ device. For a very thin spacer layer, the in-plane momentum $k_\parallel$ can be treated as conserved[26]. However, when the spacer layer becomes thicker, the coherent degree of $k_\parallel$ declines with the increasing probability of scatterings. After considering both the band structure of the GaSe spacer layer (related to the electron wave function decay) and the $k_\parallel$ scatterings (related to the coherent degree of $k_\parallel$), the bias-dependent TMR of the Fe₃GaTe₂/GaSe/Fe₃GaTe₂ MTJ device can be well explained. Under an applied voltage, electrons tunnel from the occupied states below the Fermi energy of one electrode to the empty states above the Fermi energy of the other electrode. Based on the Landauer-Buttiker (LB) formula, the spin-dependent tunneling conductance in the MTJ can be expressed as

$$G(V_b) \propto \sum_{\beta,\tau=\uparrow,\downarrow} \int T_{\beta,\tau} \times [f(E-eV_b)-f(E)]dE, \quad (1)$$

where $\beta$ and $\tau$ are the spin coefficients of the bottom and top ferromagnetic electrodes (↑,↓ represent up-spin and down-spin, respectively), $f(E)$ is the Fermi-Dirac distribution at a finite temperature, and $T_{\beta,\tau}$ is the transmission function of the MTJ. After considering both the band structure of the GaSe barrier layer and the $k_\parallel$ scatterings, the conductance of the MTJ device under parallel and antiparallel states can be written as[27,28]:

$$G_P(V_b) \propto \int (T_{\uparrow,\uparrow}+T_{\downarrow,\downarrow}) \times [f(E-eV_b)-f(E)]dE,$$

$$G_{AP}(V_b) \propto \int (T_{\uparrow,\downarrow}+T_{\downarrow,\uparrow}) \times [f(E-eV_b)-f(E)]dE,$$

$$T_{\beta,\tau} \propto \iint D_{B,\beta}(k_{\parallel B}, E-eV_b) D_{T,\tau}(k_{\parallel T}, E) e^{-2\kappa d} P(k_{\parallel B}, k_{\parallel T}) dk_{\parallel B} dk_{\parallel T},$$

$$P(k_{\parallel B}, k_{\parallel T}) = \frac{1}{2\pi\sigma^2} e^{-\frac{1}{2\sigma^2}(k_{\parallel B}-k_{\parallel T})^2}. \quad (2)$$

Here, $k_{\parallel B}$ and $k_{\parallel T}$ describe the $k_\parallel$ states of the tunneling electrons in the bottom and top electrodes, respectively. $D_{B,\beta}(k_{\parallel B}, E-eV_b)$ and $D_{T,\tau}(k_{\parallel T}, E)$ are derived from the projected density of states of the ferromagnetic electrodes [28]. The $k_\parallel$ scattering term $P(k_{\parallel B}, k_{\parallel T})$ represents the probability of a tunneling electron being scattered from $k_{\parallel B}$ state to $k_{\parallel T}$ state, which is estimated by two-dimensional Gaussian distribution. The Gaussian distribution factor $\sigma$ is calculated by a semiclassical method with $\sigma = \sigma_0\sqrt{d/\lambda}$. $\lambda$ is the mean free path of the tunneling electrons in the barrier along the perpendicular direction[29,30], and $\sigma_0$ is the Gaussian distribution factor for a single scattering, which can be estimated approximately by the Rutherford scattering model. The decay term $e^{-2\kappa d}$ describes the electron wave function decay during the tunneling process. The attenuation factor $\kappa \sim \sqrt{2m_{\parallel}(U_{\parallel} - E)/\hbar^2}$ is dependent on the band structure of the GaSe barrier layer (details see Supplementary Information Note 1) and $d$ is the barrier layer thickness. The schematic diagrams of the $k_{\parallel}$-resolved tunneling model without and with considering the GaSe barrier layer are shown in Fig. S7.

Figure 3a shows the $k_{\parallel}$-resolved conduction channels without considering the barrier layer, which are defined as $\int D_{B,\beta}(k_{\parallel}, E - eV_b) D_{T,\tau}(k_{\parallel}, E) \times [f(E - eV_b) - f(E)]dE$. With increasing the bias voltage, the distribution of the conduction channels changes differently for parallel and antiparallel states. Thus the TMR values can be influenced by the different tunneling probabilities at different $k_{\parallel}$ points due to the band structure of the barrier layer (the $e^{-2\kappa d}$ term). As shown in Fig. 3b, the conduction band of the GaSe barrier layer reaches the minimum value at Γ point, that is, electrons near Γ point meet the lowest barrier height ($U_{\parallel}$, in the attenuation factor term), thus have the highest tunneling probability. Therefore, the gathering of the conduction channels near Γ point for the parallel state will enhance the TMR value (Fig. 3a, P state, 0.2V). With increasing the bias voltage further, the conduction channels of the antiparallel state also gather towards Γ point, and even become stronger, which leads to the M-shaped bias-dependent TMR and even negative TMR values at certain bias voltage regime (Fig. 3a, AP state, 0.6 & 0.8V).

After taking the band structure of the spacer layer GaSe into account, the TMR-$V_b$ relationship of the Fe$_3$GaTe$_2$/GaSe/Fe$_3$GaTe$_2$ MTJ (mentioned in Fig. 1) is resolved from Eq. (2) and shown in Fig. 3c. The unconventional (M-shaped) TMR-$V_b$ behavior is consistent with the experimental results. As a contrast, without considering the GaSe band structure, the conventional TMR-$V_b$ relationship is observed (Fig. 3d), which verifies the important spin filter effect of the GaSe barrier layer. Therefore, we can conclude that the $k_\parallel$-resolved coherent tunneling process plays a very important role in the bias-dependent TMR relationship.

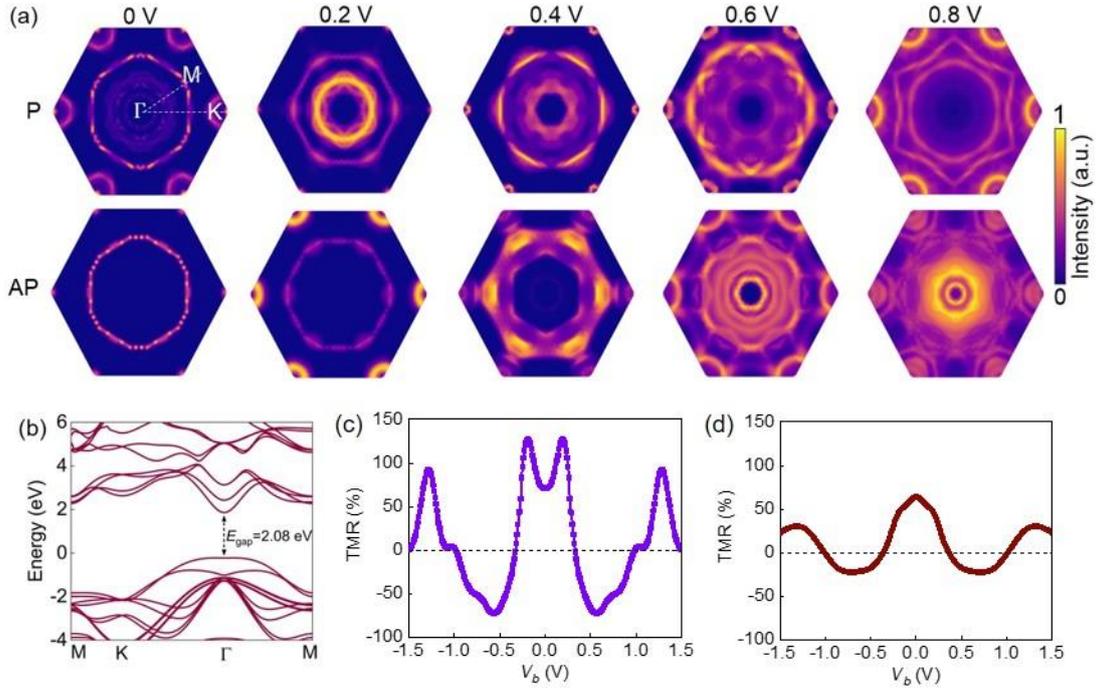

Fig. 3. Theoretical analysis of bias-dependent TMR. (a) Schematic diagram illustrating the $k_\parallel$-resolved conduction channels in the $Fe_3GaTe_2$/vacuum/$Fe_3GaTe_2$ structure at different bias voltages in both parallel and antiparallel states, neglecting the tunnel barrier layer. (b) The band structure of bulk GaSe obtained from the first-principles calculations using the hybrid functional HSE06, where the bottom of the conduction band and the top of the valence band are located at Γ. (c-d) The calculated TMR as a function of $V_b$ with (c) and without (d) considering the band structure of the GaSe barrier layer.

To further confirm the validity of our model, we modeled the TMR-$V$ relationship for a different MTJ, based on $Fe_3GeTe_2$/GaSe/$Fe_3GeTe_2$[22]. For the $Fe_3GeTe_2$ electrode, the conduction channel distribution in the P state does not show a more obvious aggregation towards the Γ point than in the AP state with increasing bias voltage. Thus, the calculated TMR begins to decrease from zero bias even if we consider both the coherent degree of $k_\parallel$ and the decay of electron wave function, in contrast to the Fe$_3$GaTe$_2$/GaSe/Fe$_3$GaTe$_2$ MTJ. As shown in Fig. S8, the magnitude of TMR near zero bias and TMR sign transition point are more consistent with the experimental observations, and significantly better than the previous theoretical predictions[22].

## 3. Conclusion

In summary, we observed an unconventional bias-dependent TMR effect in the all-vdW Fe$_3$GaTe$_2$/GaSe/Fe$_3$GaTe$_2$ MTJ devices. With increasing the bias voltage in both positive and negative directions, an unconventional M-shaped bias-dependent TMR was observed with a maximum TMR of 107% under a bias of 0.5 V at 10 K, which cannot be explained using a traditional tunneling model. Thus, to reveal the underlying physics of the unconventional bias-dependent TMR, we propose a $k_\parallel$-resolved tunneling model that incorporates both the coherent degree of $k_\parallel$ and the decay of the electron wave function induced by the semiconductor spacer layer. Using our model, not only the unconventional bias-dependent TMR observed in this work, but also the conventional bias-dependent TMR can be elucidated. Our work lays a solid foundation for understanding the physical origin of the bias-dependent TMR for semiconductor-based all-vdW MTJs and developing high-performance semiconductor spintronic devices.

## 4. Experimental Section

*MTJ device Fabrication*: The high-quality vdW single-crystal Fe$_3$GaTe$_2$ was grown using the self-flux method. GaSe was purchased from 2D Semiconductors, while WSe$_2$ and hBN were purchased from HQ Graphene. The metal contact electrodes were pre-patterned by standard photolithography, and Cr (5 nm)/Au (45 nm) layers were deposited on the Si/300-nm-SiO$_2$ substrate by using the ultrahigh vacuum magnetron sputtering system, followed by a lift-off process. The MTJ devices were fabricated using a polydimethylsiloxane (PDMS)-assisted 2D dry-transfer method. First, blue tape is used to thin the bulk of the vdW material. Then, the blue tape is attached to the

PDMS/glass sheet to transfer the vdW nanoflakes onto the PDMS. After that, under an optical microscope, appropriate shapes and thicknesses of the vdW nanoflakes are selected according to their colors. The nanoflakes are then transferred to a specific location on the substrates by using a position-controllable dry-transfer method. In MTJ devices, the thickness of the bottom layer of $Fe_3GaTe_2$ is approximately 8-12 nm, the thickness of the top layer of $Fe_3GaTe_2$ is about 15-20 nm, and the semiconductor layer is about 5-10 nm thick. The MTJ devices are covered with a protective layer of 30-nm-hBN to prevent oxidation. The MTJ devices are finally baked at 120° for 10 minutes after the transfer process to ensure closer contact between the interfaces of different vdW materials. The entire fabrication processes are carried out within a glovebox, with an internal water and oxygen content of less than 0.1 ppm to prevent degradation of the material quality.

*Characterization*: The thickness of the vdW nanoflakes was characterized by AFM (Bruker Multimode 8).

*Magnetoresistance Measurements*: The electromagnetic transport properties of the device were investigated within a Model CRX-VF Cryogenic Probe Station (Lake Shore Cryotronics, Inc.) with a vertical magnetic field of ±2.5 T and a variable temperature range from 10 K to 500 K. The resistance of the device was monitored by a semiconductor parameter analyzer (Agilent B1500A).

*Theoretical Calculation*: In the study, the first principles calculation was performed by using the Vienna ab initio simulation package (VASP)[31]. The projector augmented wave (PAW) potentials were applied to the elements[32], and the generalized gradient approximation (GGA) exchange-correlation function was employed[33]. The DFT-D3 correction was used for vdW interaction[34]. The Brillouin zone sampling was performed by using the gamma-centered k-meshes (30×30×6) for the accurate spin-resolved electronic structure calculation.

**Acknowledgements**

This work was supported by the National Key Research and Development Program of China (Grant Nos. 2022YFA1405100 and 2022YFE0134600), the Beijing Natural Science Foundation Key Program (Grant No. Z220005), the National Natural Science Foundation of China (Grant Nos. 12241405, 12174384, 52272152, and 12404146). H. W., S. Z., and K. W. thank the very fruitful discussions with Prof. Igor Žutić, and Prof. Kirill D. Belashchenko for theoretical model.

**Author contributions**

K.W. conceived the work. W.Z. fabricated the devices, W.Z., S.Z., and S.X. performed the experiments. W.Z., H. W., S.Z., Q.C., and K.W. analyzed the data. H.W., S.Z., Q.C., M.Y., L.W.W., and K.W. carried out the modeling. H.W., S.Z., and Q.C. performed DFT calculations. G.Z., H.W., and H.C. provided the $Fe_3GaTe_2$ bulk crystals and conducted the preliminary studies of $Fe_3GaTe_2$. W.Z., H.W., S.Z., Q.C., Y.H., J.Z., L.Z., A.P., and K.W. wrote the manuscript. All authors discussed the results and commented on the manuscript.

**Conflict of Interest**

The authors declare no conflict of interest.

**Data Availability Statement**

The data that support the findings of this study are available from the corresponding author upon reasonable request.

**Code availability**

The codes that support the theoretical part of this study are also available from the corresponding author upon reasonable request.

**Additional information**

**Supplementary information** is available for this paper.

**Correspondence and requests for materials** should be addressed to H.C. and K.W.

**Reprints and permissions information** is available at www.nature.com/reprints.